\documentclass[12pt]{article}
\usepackage{amsmath}
\usepackage{amssymb}
\usepackage{amsthm}
\usepackage{graphicx}
\newcommand{\pa}{\partial}
\newcommand{\bm}[1]{\textnormal{\mathversion{bold}$#1$}}
\newcommand{\defeq}{\stackrel{\mathrm{def}}{=}}
\newcommand{\IC}{\mathbb{C}}
\newcommand{\IZ}{\mathbb{Z}}
\newcommand{\slt}{\mathfrak{sl}^{\mathrm{tor}}}
\newcommand{\slh}{\widehat{\mathfrak{sl}}}
\newcommand{\ii}{\mathrm{i}}
\newcommand{\dd}{\mathrm{d}}
\newcommand{\OmegaTor}{\Omega^{\mathrm{tor}}}

\newtheorem{proposition}{Proposition}
\newtheorem{lemma}{Lemma}

\title{
Differential-difference system related to \\
toroidal Lie algebra}
\author{
Saburo Kakei
\footnote{Present address: 
Department of Mathematics, Rikkyo University, 
Nishi-ikebukuro 3-34-1, Toshima-ku, Tokyo 171-8501, Japan. 
E-mail: kakei@rkmath.rikkyo.ac.jp}\\
{\normalsize Department of Mathematical Sciences, }\\
{\normalsize School of Science and Engineering, Waseda University}\\
{\normalsize Ohkubo 3-4-1, Shinjuku-ku, Tokyo 169-8555, Japan}\\[3mm]%
Yasuhiro Ohta\\
{\normalsize Department of Applied Mathematics, }\\
{\normalsize Faculty of Engineering, Hiroshima University}\\
{\normalsize Kagamiyama 1-4-1, Higashi-hiroshima 739-8527, Japan}\\
{\normalsize E-mail: ohta@kurims.kyoto-u.ac.jp}}
\date{}
\begin{document}
\maketitle
\begin{abstract}
We present a novel differential-difference system in (2+1)-dimensional 
space-time (one discrete, two continuum), arisen from the 
Bogoyavlensky's (2+1)-dimensional KdV hierarchy. Our method is based on 
the bilinear identity of the hierarchy, which is related to the vertex 
operator representation of the toroidal Lie algebra $\slt_2$.
\end{abstract}

\section{Introduction and main results}
Multi-dimensional generalization of classical soliton equations 
has been one of the most exciting topic in the field of integrable
systems. 
Among other things, Calogero \cite{Cal} proposed an interesting 
example that is a (2+1)-dimensional extension of the Korteweg-de Vries
equation, 
\begin{equation}
u_t=\frac{1}{4}u_{xxy} + uu_y + \frac{1}{2}u_x \int^x u_y \dd x.
\label{2dKdV}
\end{equation}
Yu et al.~\cite{YTSF} 
obtained multi-soliton solutions of the (2+1)-dimensional KdV
equation \eqref{2dKdV} by using the Hirota's bilinear method. 
Let us consider the following Hirota-type equations, 
\begin{align}
 \left(D_{x}^4-4D_{x}D_{t'}\right)&\tau\cdot\tau=0,\label{Bil_KdV}\\
 \left(D_{y}D_{x}^3+2D_{y}D_{t'}-6D_{t}D_{x}\right)
 & \tau\cdot\tau=0\label{Bil_2+1d}, 
\end{align}
where we have used the $D$-operators of Hirota defined as 
\[
\begin{aligned}
D_x D_y &\ldots f(x,y,\ldots)\cdot g(x,y,\ldots)\\
 &\defeq\left. \pa_s\pa_t 
f(x+s,y+t,\ldots)g(x-s,y-t,\ldots)\right|_{s,t,\ldots=0} .
\end{aligned}
\]
We remark that 
we have introduced auxiliary variables $t'$ that is a hidden
parameter in \eqref{2dKdV}. 
If we set $u = 2(\log\tau)_{xx}$ and use \eqref{Bil_KdV} to eliminate
$\pa_{t'}$, then one can show that $u=u(x,y,t)$ solves \eqref{2dKdV}. 

Bogoyavlensky \cite{Bog} showed that there is a hierarchy of
higher-order integrable equations associated with \eqref{2dKdV}. 
In the paper \cite{IT}, Ikeda and Takasaki generalized the
Bogoyavlensky's hierarchy from the viewpoint of the Sato's theory 
of the KP hierarchy \cite{Sato,SS,DJKM,JM}, and 
discussed the relationship to toroidal Lie algebras. 
We note that the relation between integrable hierarchy and toroidal
algebras has been discussed also by Billig \cite{Bi}, Iohara, Saito and
Wakimoto \cite{ISW} by using vertex operator representations. 

In the present paper, we propose the following
differential-difference system that has the same symmetry: 
\begin{eqnarray}
\partial_t u_k &=& \Delta_{-k}\left(
\frac{\partial_x u_{k+1}}{1-\exp(-u_{k+1}-u_k)}
\right.\nonumber\\
&&\qquad\quad\left.
-\frac{\partial_x u_k}{1-\exp(u_{k+1}+u_k)}
-\frac{1+\exp(u_{k+1}+u_k)}{1-\exp(u_{k+1}+u_k)}v_k
\right), \label{DDeq1}\\
\Delta_{-k} v_k &=&
\frac{\partial_x u_{k+1}}{u_{k+1}}
+\frac{\partial_x (u_{k+1}+u_k)}{1-\exp(u_{k+1}+u_k)}
\nonumber\\
&&\qquad\quad
+\frac{\partial_x u_k}{u_k}
+\frac{\partial_x (u_k+u_{k-1})}{1-\exp(u_k+u_{k-1})}, 
\label{DDeq2}
\end{eqnarray}
where $\Delta_{-k}$ denotes the backward difference operator 
$\Delta_{-k} \defeq 1 - \exp(-\partial_k)$ 
($\Delta_{-k} u_k = u_k-u_{k-1}$). 
We also show that this system has soliton-type solutions. 

\section{Lie algebraic derivation of bilinear identity}
Here we briefly review the Lie algebraic derivation of the bilinear
identity of the Bogoyavlensky's hierarchy \cite{IT}, 
which is a generating function of Hirota-type differential equations. 
We remark that the Lie algebra considered in \cite{IT} is bigger than
that of present article. 
Here we don't include the derivations to $\slt_2$ since those are not
needed for our purpose. 
Owing to this difference, the proof given below may be simpler than that
of \cite{IT}. 

The 2-toroidal Lie algebra $\slt_2$ \cite{Kass,MEY} is the universal
central extension of the double loop algebra
$\mathfrak{sl}_2\otimes\IC[s,s^{-1},t,t^{-1}]$, 
while the affine Lie algebra $\slh_2$ is the central extension of 
$\mathfrak{sl}_2\otimes\IC[t,t^{-1}]$. 
Let $A$ be the ring of Laurent polynomials of two variables $s$ and $t$. 
As a vector space, $\slt_2$ is isomorphic to 
$\mathfrak{sl}_2\otimes\IC[s,s^{-1},t,t^{-1}] \oplus \Omega_A/\dd A$,
where $\Omega_A$ denotes the module of K\"{a}hler differentials of $A$
defined with the canonical derivation $\dd:A\rightarrow \Omega_A$.
We define the Lie algebra structure of $\slt_2$ by
\begin{eqnarray}
&& [ x\otimes a, y\otimes b ] = 
[x,y]\otimes ab + (x|y) \overline{(\dd a)b}
\quad (x, y\in\mathfrak{sl}_2,\; a,b\in A), \label{CRofsl2tor}\\
&& [\slt_2, \Omega_A/\dd A] = 0, 
\end{eqnarray}
where $(x|y)$ denotes the Killing form and 
$\overline{\cdot}:\Omega_A\rightarrow \Omega_A/\dd A$ 
the canonical projection. 

In terms of generating series $X_{m}(z)$ ($X=E,F,H$, $m\in\IZ$), 
$K_{m}^s(z)$ and $K_{m}^t(z)$, defined by 
\begin{eqnarray*}
 X_{m}(z) &\defeq&
   \sum_{n\in\IZ}X\otimes s^n t^m \cdot z^{-n-1},\\
 K_{m}^s(z) &\defeq&
   \sum_{n\in\IZ}\overline{s^n t^m \,\dd\log s}\cdot z^{-n},\\
 K_{m}^t(z) &\defeq&
   \sum_{n\in\IZ}\overline{s^n t^m \,\dd\log t}\cdot z^{-n-1},  
\end{eqnarray*}
the relation \eqref{CRofsl2tor} can be expressed as 
\begin{eqnarray}
\lefteqn{X_n(z)Y_n(w) \;=\;
\frac{1}{z-w}[X,Y]_{m+n}(w)} \nonumber\\
&& \;+\; \frac{1}{(z-w)^2}(X|Y)K_{m+n}^s(w) 
   \;+\; \frac{m}{z-w}(X|Y)K_{m+n}^{t}(w) \label{torOPE}\\
&& \;+\; \mbox{ regular as }\;z\rightarrow w. \nonumber
\end{eqnarray}

There exists a class of representations of $\slt_2$, which comes
directly from that of $\slh_2$. 
We consider the space of polynomials, 
\[
F_y \defeq 
\IC[y_j,\,j\in\IZ]\otimes\IC[\exp(\pm y_0)], 
\]
and define the generating series $\varphi(z)$ and $V_{m}(y;z)$ by 
\[
 \varphi(z)\defeq\sum_{n\in\IZ}n y_n z^{n-1}, \qquad
 V_{m}(y;z)\defeq
 \exp\left[ m\sum_{n\in\IZ}y_nz^n \right]. 
\]
\begin{proposition} (cf. \cite{ISW,BB})
\label{prop:tor_ext}
Let $(V,\pi)$ be a representation of $\slh_2$ 
such that $\overline{\dd\log s}\mapsto c\cdot\mathrm{id}_V$ for $c\in\IC$.
Then we can define the representation $\pi^{\mathrm{tor}}$ of $\slt_2$ on
$V\otimes F_y$ such that 
\begin{eqnarray*}
 X_{m}(z) &\mapsto& X^\pi(z)\otimes V_{m}(z), \\
 K_{m}^s(z) &\mapsto& c\cdot\mathrm{id}_V \otimes V_{m}(z), \\
 K_{m}^t(z) &\mapsto&
    c\cdot\mathrm{id}_V \otimes \varphi(z)V_{m}(z),
\end{eqnarray*}
where $X=E,F,H$, $m\in\IZ$ and
 $X^\pi(z)\defeq\sum_{n\in\IZ}\pi(X\otimes s^n)z^{-n-1}$.
\end{proposition}
\begin{proof}
Using the operator product expansion for $\slh_2$, 
\begin{eqnarray*}
X(z)Y(w) &=&  \frac{1}{z-w}[X,Y](w) + \frac{1}{(z-w)^2}(X|Y)K\\
&& \;+\; \mbox{ regular as }\;z\rightarrow w, 
\end{eqnarray*}
and the property $V_{m}(z)V_{n}(z)=V_{m+n}(z)$, 
it is straightforward to show that $X_{m}(z)$ above satisfies \eqref{torOPE}. 
The remaining relations can be checked by direct calculations.
\end{proof}

To see the relationship to soliton theory, we shall consider the 
representation of $\slh_2$ on the fermionic Fock space \cite{DJKM,JM}.
Let $\psi_j$, $\psi_j^*$ ($j\in\IZ$) be free fermions with the 
canonical anti-commutation relation. 
In terms of the generating series defined as
\[
\psi(\lambda)=\sum_{n\in\IZ}\psi_n\lambda^n, \qquad
\psi^*(\lambda)=\sum_{n\in\IZ}\psi^*_n\lambda^{-n}, 
\]
the canonical anti-commutation relation is written as 
\begin{equation}
\label{CaCRlambda}
\left[\psi(\lambda), \psi^*(\mu)\right]_+
= \delta(\lambda/\mu), \qquad 
\left[\psi(\lambda), \psi(\mu)\right]_+
= \left[\psi^*(\lambda), \psi^*(\mu)\right]_+=0, 
\end{equation}
where $\delta(\lambda)\defeq\sum_{n\in\IZ}\lambda^n$ is the 
formal delta-function. 

Consider the fermionic Fock space ${\cal F}$ with the vacuum vector
$|\mathrm{vac}\rangle$ satisfying
\[
 \psi_j|\mathrm{vac}\rangle = 0 \;\mbox{ for }\; j<0, \qquad
 \psi^*_j|\mathrm{vac}\rangle = 0 \;\mbox{ for }\; j\ge 0, 
\]
and the dual Fock space ${\cal F}^*$ with the dual vacuum vector
$\langle\mathrm{vac}|$ satisfying
\begin{eqnarray*}
&& \langle\mathrm{vac}| \psi_j = 0 \;\mbox{ for }\; j\ge 0, \qquad
\langle\mathrm{vac}| \psi^*_j = 0 \;\mbox{ for }\; j< 0, \\
&& \langle\mathrm{vac}|\mathrm{vac}\rangle = 1. 
\end{eqnarray*}
As mentioned in \cite{DJKM,JM}, a level-1 representation of $\slh_2$ is given
by the elements, 
\[
 :\! \psi(\lambda)\psi^*(-\lambda) \! : \;=\;
\sum_{j,n\in\IZ}(-1)^j
:\! \psi_{j+n}\psi^*_j \! : \lambda^n , 
\]
where $:\!\cdot\! :$ denotes the fermionic normal ordering, 
$:\!\psi_i \psi^*_j\!:\, \defeq \psi_i \psi^*_j- 
\langle\mathrm{vac}|\psi_i \psi^*_j|\mathrm{vac}\rangle$. 
Applying Proposition \ref{prop:tor_ext}, we can construct a 
representation of $\slt_2$ on the space 
${\cal F}_y\defeq {\cal F} \otimes F_y$ with the vacuum vector 
$|\mathrm{vac}\rangle^{\mathrm{tor}}\defeq |\mathrm{vac}\rangle\otimes 1$. 

We now introduce the following operator acting on 
${\cal F}_y\otimes{\cal F}_{y'}$: 
\[
\OmegaTor \defeq 
\sum_{m\in\IZ}\oint\frac{\dd\lambda}{2\pi\ii\lambda}\,
\psi(\lambda)V_m(\lambda;y)\otimes
\psi^*(\lambda)V_{-m}(\lambda;y'). 
\]
Using the anti-commutation relation \eqref{CaCRlambda} and the relation 
$V_{m}(y;\lambda)V_{n}(y;\lambda)=V_{m+n}(y;\lambda)$, we can obtain the
following identity by direct calculations: 
\[
 [ \OmegaTor,\, 
   \psi(p)\psi^*(p)V_n(y;p)\otimes 1
 + 1\otimes\psi(p)\psi^*(p)V_n(y';p)] = 0, 
\]
which means the action of $\slt_2$ on ${\cal F}_y\otimes{\cal F}_{y'}$
commutes with $\OmegaTor$. 
Then it is straightforward to show that
\begin{equation}
\label{BIinF}
\OmegaTor \left( g|\mathrm{vac}\rangle^{\mathrm{tor}}
\otimes g|\mathrm{vac}\rangle^{\mathrm{tor}} \right)=0
\end{equation}
for $g=\exp(X)$, $X\in\slt_2$. 

To rewrite \eqref{BIinF} into bosonic language, we prepare two lemmas: 
\begin{lemma}[``Boson-Fermion correspondence''] \cite{DJKM,JM}
\label{Lem:BFcorr}
For any $|\nu\rangle\in{\cal F}$, we have the following formulas, 
\[
\begin{aligned}
\langle\mathrm{vac}|\psi^*_0\exp(H(\bm{x}))\psi(\lambda)|\nu\rangle
&= \exp(\xi(\bm{x},\lambda))
    \langle\mathrm{vac}|\exp(H(\bm{x}-[\lambda^{-1}]))|\nu\rangle,\\
\langle\mathrm{vac}|\psi_{-1}\exp(H(\bm{x}))\psi^*(\lambda)|\nu\rangle
&= \lambda\exp(-\xi(\bm{x},\lambda))
    \langle\mathrm{vac}|\exp(H(\bm{x}+[\lambda^{-1}]))|\nu\rangle,\\
\end{aligned}
\]
where we have used the following notation, 
\[
\begin{aligned}
\bm{x}=(&x_1, x_3,\ldots), \qquad 
H(\bm{x})\defeq \sum_{n=1}^{\infty}\sum_{j\in\IZ}x_n \psi_j\psi_{n+j}^*, \\
\xi(\bm{x},\lambda) &\defeq \sum_{n=1}^{\infty}x_n \lambda^n, \qquad
[\lambda^{-1}] \defeq \left(\,
1/\lambda, 1/2\lambda^2, 1/3\lambda^3,\ldots \,\right). 
\end{aligned}
\]
\end{lemma}
\begin{lemma} \cite{Bi,ISW}
\label{Lem:Billig}
Let $P(n)=\sum_{j\geq 0}n^j P_j$, where $P_j\in$ Diff$(z)$ are 
differential operators that may not depend on $z$. 
If
\[
\sum_{n\in\IZ}z^n P(n)g(z)=0 
\]
for some formal series $g(z)=\sum_j g_jz^j$, then 
\[
\left.P(\epsilon-z\pa_z)g(z)\right|_{z=1}=0
\]
as a polynomial in $\epsilon$. 
\end{lemma}

Define the $\tau$-function as 
\[
\tau(\bm{x},\bm{y}) \defeq 
\,^{\mathrm{tor}}\langle\mathrm{vac}| \exp(H(\bm{x}))
g|\mathrm{vac}\rangle^{\mathrm{tor}}. 
\]
{}From the relation \eqref{BIinF}, together with Lemma
\ref{Lem:BFcorr} and Lemma \ref{Lem:Billig}, we have the 
following bilinear identity: 
\begin{align}
\oint\frac{\dd\lambda}{2\pi\ii}\;\exp(\xi(\bm{x}-\bm{x}',\lambda))
& \tau(\bm{x}-[\lambda^{-1}],
       y_0+\eta(\check{\bm{b}},\lambda^2),\check{\bm{y}}-\check{\bm{b}})
\nonumber\\
\times & \tau(\bm{x}'+[\lambda^{-1}],
y_0-\eta(\check{\bm{b}},\lambda^2),\check{\bm{y}}+\check{\bm{b}}) 
= 0,  \label{BI}
\end{align}
where $\check{\bm{y}}\defeq (y_2,y_4,\ldots)$ and 
$\eta(\check{\bm{b}},\lambda^2) \defeq \sum_{n=1}^{\infty}b_{2n}\lambda^{2n}$. 

Expanding \eqref{BI}, we can obtain Hirota-type differential equations
including \eqref{Bil_KdV}, \eqref{Bil_2+1d} 
($x_1=x$, $x_3=t'$, $y_0=y$, $y_2=t$). 
In this sense, the bilinear identity \eqref{BI} is a 
generating function of Hirota-type differential equations 
of the Bogoyavlensky's hierarchy.  
The $N$-soliton solution of \eqref{BI} is obtained as follows \cite{IT}: 
\begin{eqnarray}
\tau_N(\bm{x},\bm{y}) &=& \sum_{l=0}^{N}\sum_{j_1<\cdots<j_l}
 c_{j_1\cdots j_l} \prod_{m=1}^{l}a_{j_m}
 \exp(\eta_{j_m}(\bm{x},\bm{y})), \label{Nsoliton}\\
\eta_j(\bm{x},\bm{y}) &\defeq& 
\sum_{n=1}^{\infty}2p_j^{2n-1}x_{2n-1}+
\sum_{n=1}^{\infty}r_jp_j^{2n}y_{2n}, \nonumber\\
c_{j_1\cdots j_l} &\defeq& \prod_{1\leq m < n \leq l}
 \frac{(p_{j_m}-p_{j_n})^2}{(p_{j_m}+p_{j_n})^2}. \nonumber
\end{eqnarray}

\section{Derivation of the differential-difference system}

We now apply the Miwa transformation \cite{JM,M}, 
\begin{equation}
\label{MiwaTr}
\begin{array}{rcl}
\bm{x}' &=& l [\alpha] + m [\beta] + n [\gamma], \\
\bm{x} &=& (l+1) [\alpha] + (m+1) [\beta] + (n+1) [\gamma], 
\end{array}
\end{equation}
to the bilinear identity \eqref{BI}. 
Here we have used the notation 
$l [\alpha] = (l\alpha,l\alpha^2/2,l\alpha^3/3,\ldots)$. 
We first consider the case 
$\check{\bm{b}}=(b_2,b_4,\ldots)=\bm{0}$. 
In this case, the bilinear identity \eqref{BI} is reduced to that of 
the ordinary KP hierarchy. Thus we have the Hirota-Miwa equation 
(or the discrete KP equation), 
\begin{equation}
\begin{aligned}
\alpha(\beta-\gamma)&\;\tau(l+1,m,n;\bm{y})\tau(l,m+1,n+1;\bm{y})\\
+\beta(\gamma-\alpha)&\;\tau(l,m+1,n;\bm{y})\tau(l+1,m,n+1;\bm{y})\\
+\gamma(\alpha-\beta)&\;\tau(l,m,n+1;\bm{y})\tau(l+1,m+1,n;\bm{y})
=0
\end{aligned}
\label{HM}
\end{equation}
where $\tau(l,m,n;\bm{y})$ denotes 
\[
\tau(l,m,n;\bm{y})=\tau(\bm{x}=l[\alpha]+m[\beta]+n[\gamma],\bm{y}). 
\]

We then consider the time-evolutions with respect to $y_0$, $y_2$. 
Collecting the coefficient of $b_2$ in the bilinear identity \eqref{BI}, 
we have
\[
\begin{aligned}
\oint\frac{\dd\lambda}{2\pi\ii}\;&
\exp(\xi(\bm{x}-\bm{x}',\lambda))
\left(D_{y_2}-\lambda^2 D_{y_0}\right)\\
\times &\tau(\bm{x}'+[\lambda^{-1}],\bm{y})\cdot
\tau(\bm{x}-[\lambda^{-1}],\bm{y})= 0 .
\end{aligned}
\]
Applying the Miwa transformation \eqref{MiwaTr}, we obtain 
\begin{equation}
\begin{aligned}
\,&\alpha^2\beta\gamma(\beta-\gamma)D_{y_2}
\tau(l+1,m,n;\bm{y})\cdot\tau(l,m+1,n+1;\bm{y})\\
\,&+\alpha\beta^2\gamma(\gamma-\alpha)D_{y_2}
\tau(l,m+1,n;\bm{y})\cdot\tau(l+1,m,n+1;\bm{y})\\
\,&+\alpha\beta\gamma^2(\alpha-\beta)D_{y_2}
\tau(l,m,n+1;\bm{y})\cdot\tau(l+1,m+1,n;\bm{y})\\
= & \;\beta\gamma(\beta-\gamma)D_{y_0}
\tau(l+1,m,n;\bm{y})\cdot\tau(l,m+1,n+1;\bm{y})\\
\,&+\gamma\alpha(\gamma-\alpha)D_{y_0}
\tau(l,m+1,n;\bm{y})\cdot\tau(l+1,m,n+1;\bm{y})\\
\,&+\alpha\beta(\alpha-\beta)D_{y_0}
\tau(l,m,n+1;\bm{y})\cdot\tau(l+1,m+1,n;\bm{y})\\
\,&-(\alpha-\beta)(\beta-\gamma)(\gamma-\alpha)\\
\,&\qquad\times D_{y_0}
   \tau(l,m,n;\bm{y})\cdot\tau(l+1,m+1,n+1;\bm{y})
\end{aligned}
\label{semidisc}
\end{equation}

We further impose the condition $\beta=\gamma$. Then the $\tau$-function 
$\tau(l,m,n;\bm{y})$ depends only on $k\defeq m-n$, $l$ and $\bm{y}$. 
In this sense, we rewrite 
\[
 \tau(l,m,n;\bm{y}) \;\;\rightarrow\;\; \tau(l,k;\bm{y}) .
\]
Under this condition, the equations \eqref{HM} and \eqref{semidisc} are 
reduced to
\begin{align}
\,&2\alpha\;\tau(l+1,k;\bm{y})\tau(l,k;\bm{y})\nonumber\\
-&(\alpha+\beta)\;\tau(l,k+1;\bm{y})\tau(l+1,k-1;\bm{y})\nonumber\\
-&(\alpha-\beta)\;\tau(l,k-1;\bm{y})\tau(l+1,k+1;\bm{y})=0,
\label{DifDif1}\\[2mm]
\alpha\beta^2 D_{y_2}&\left(
2\alpha\;\tau(l+1,k;\bm{y})\cdot\tau(l,k;\bm{y})
\right.\nonumber\\
\,&\quad -(\alpha+\beta)\;
\tau(l,k+1;\bm{y})\cdot\tau(l+1,k-1;\bm{y})\nonumber\\
\,&\quad \left. -(\alpha-\beta)\;
\tau(l,k-1;\bm{y})\cdot\tau(l+1,k+1;\bm{y})\right)\nonumber\\
=D_{y_0}&\left(2(2\beta^2-\alpha^2)\;
\tau(l+1,k;\bm{y})\cdot\tau(l,k;\bm{y})
\right.\nonumber\\
\,&\quad -\alpha(\alpha+\beta)\;
\tau(l,k+1;\bm{y})\cdot\tau(l+1,k-1;\bm{y})\nonumber\\
\,&\quad \left. -\alpha(\alpha-\beta)\;
\tau(l,k-1;\bm{y})\cdot\tau(l+1,k+1;\bm{y})\right). 
\label{DifDif2}
\end{align}
Furthermore, we can construct $N$-soliton solution by applying
\eqref{MiwaTr} to \eqref{Nsoliton}: 
\begin{eqnarray}
\tau_N(l,k;y_0,y_2) &=& \sum_{l=0}^{N}\sum_{j_1<\cdots<j_l}
 c_{j_1\cdots j_l} \prod_{m=1}^{l}\phi_{j_m}(l,k;y_0,y_2), 
\label{NsolDD}\\
\phi_{j}(l,k;y_0,y_2) &\defeq& a_{j}
 \exp(r_{j}y_0+r_{j}p_{j}^2 y_2)
 \left(\frac{1+p_{j}\alpha}{1-p_{j}\alpha}\right)^l
 \left(\frac{1+p_{j}\beta}{1-p_{j}\beta}\right)^k, \nonumber
\end{eqnarray}
where $c_{j_1\cdots j_l}$ is the same as the continuum one
\eqref{Nsoliton}. 
We remark that 
the $N$-soliton $\tau$-function can be written as Wronskian determinant. 
Using the determinant expression, we can show that both \eqref{DifDif1} 
and  \eqref{DifDif2} are reduced to the Pl\"ucker relations. 

Introducing the variables as 
\begin{eqnarray*}
&&\partial_t \defeq \frac{2}{\alpha^2}\partial_{y_0}-2\partial_{y_2}, \qquad
\partial_x \defeq \partial_{y_2}-\frac{1}{\beta^2}\partial_{y_0}, \\
&&u_k(t,x) \defeq \log\!\left[
\left(\frac{\beta-\alpha}{\beta+\alpha}\right)^{1/2}
\frac{\tau(l+1,k+1)\tau(l,k)}{\tau(l,k+1)\tau(l+1,k)} \right], \\
&&v_k(t,x) \defeq \partial_x \log\frac{\tau(l,k+2)}{\tau(l,k)}, 
\end{eqnarray*}
we have the differential-difference equations 
\eqref{DDeq1} and \eqref{DDeq2}, 
which have the $N$-soliton solution corresponding to the $\tau$-function 
\eqref{NsolDD}. 

\section{Concluding remarks}
\begin{figure}[htbp]
\begin{center}
\includegraphics{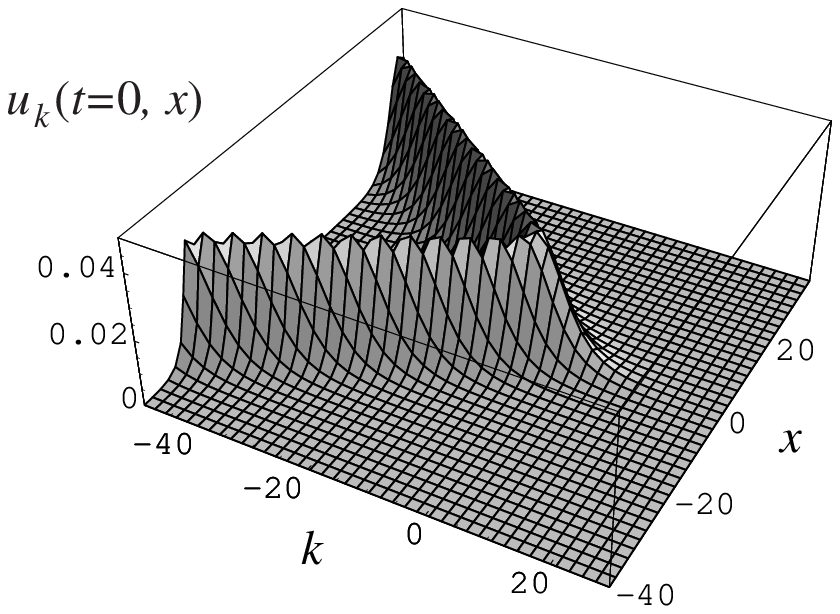}
\includegraphics{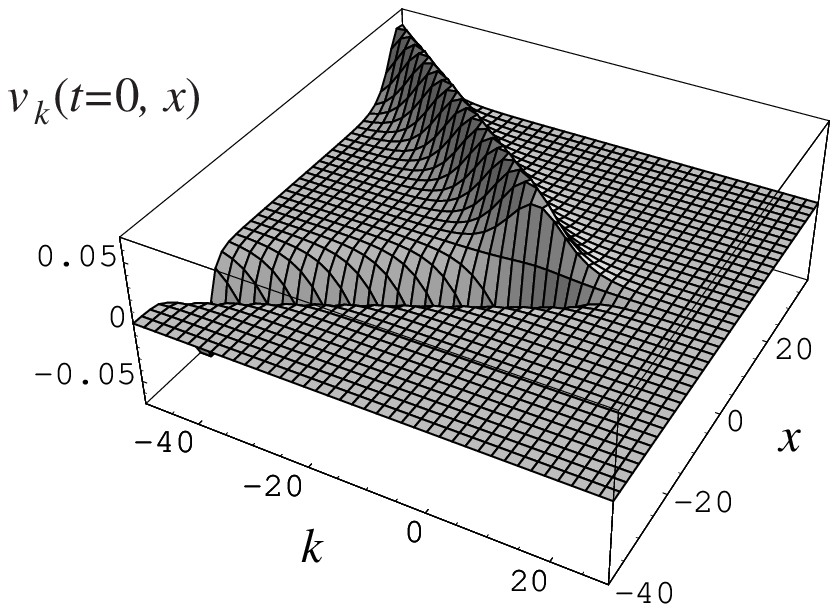}
\caption{Special case of the two-soliton solutions (``V-soliton'') with 
$p_1 = p_2$ ($p_1=p_2=0.3$, $r_1=0.15$, $r_2=-0.1$, $a_1=a_2=1$, 
$\alpha=0.8$, $\beta=0.5$).}
\label{fig:Vsoliton}
\end{center}
\end{figure}

In this paper, we have introduced the differential-difference system 
\eqref{DDeq1} and \eqref{DDeq2}, which is related to 
toroidal Lie algebra $\slt_2$. 
Since the symmetry of the toroidal Lie algebra allows us to introduce
extra parameters of wave numbers in the soliton solution 
(i.e., $r_j$ in \eqref{Nsoliton} and \eqref{NsolDD}), 
it might be possible to construct some interesting solutions.
In particular, we can obtain a class of traveling-wave solutions that 
has the shape of the character ``V'' (Figure \ref{fig:Vsoliton}), which 
is a special case of the two-soliton solutions. 
The existence of the V-soliton type solution is 
one of the features of this class of equations.

We note that there exist solutions of the same shape for the
(2+1)-dimensional KdV equation \eqref{2dKdV}, and for a
(2+1)-dimensional generalization of the nonlinear Schr\"odinger (NLS) 
equation \cite{St} that also has the symmetry of the toroidal Lie
algebra $\slt_2$ \cite{KIT}.
We also remark that Oikawa et al. \cite{OOF} discussed the
propagation of the V-soliton in a two-layer fluid, which is 
governed by a equation similar to the (2+1)-dimensional NLS equation. 

\section*{Acknowledgements}
The authors acknowledge Doctor Takeshi Ikeda and Professor 
Kanehisa Takasaki for discussions. 
The first author is partially supported by the 
Grant-in-Aid for Scientific Research (No.~12740115) from 
the Ministry of Education, Culture, Sports, Science and Technology.

\newpage
\ \\

\vspace{3cm}

\newpage
\ \\

\vspace{3cm}

\begin{figure}[htbp]
\begin{center}
\includegraphics{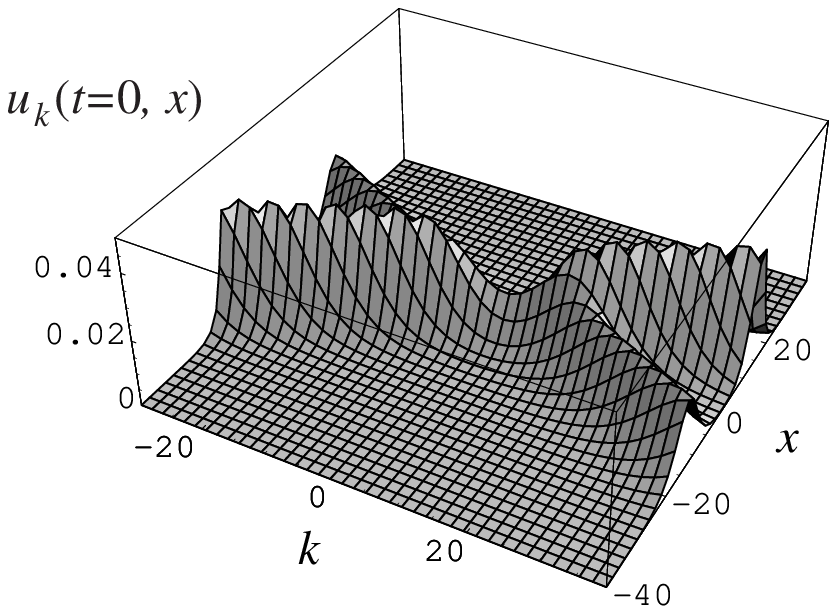}
\includegraphics{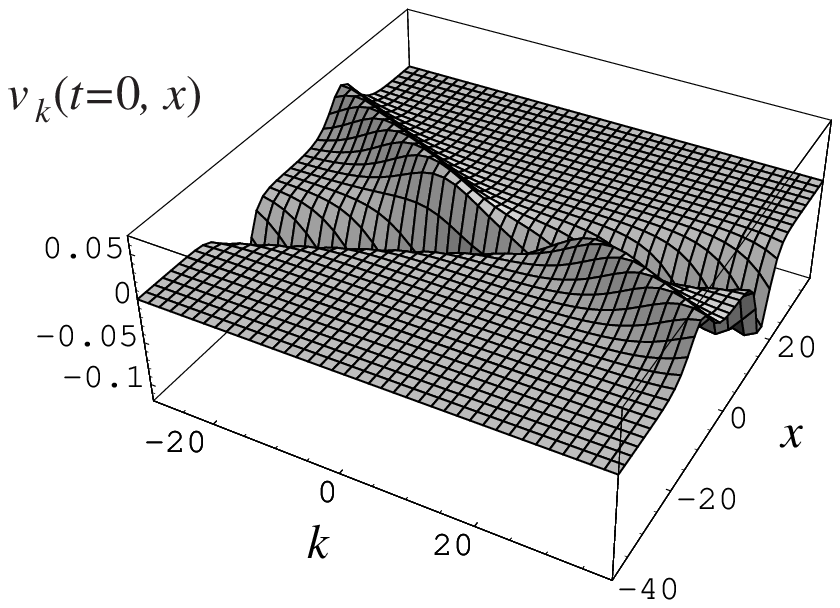}
\caption{Two-soliton solutions with $p_1\neq p_2$
($p_1=0.3$, $p_2=0.23$, $r_1=0.15$, $r_2=-0.1$, $a_1=a_2=1$, 
$\alpha=0.8$, $\beta=0.5$).} 
\label{fig:2soliton}
\end{center}
\end{figure}

\end{document}